\documentclass[journal]{IEEEtran}
\usepackage{cite}
\usepackage{amsmath}
\usepackage{upgreek}
\usepackage[pdftex]{graphicx}
\usepackage{array}
\usepackage{url}
\usepackage{color}

\usepackage[utf8]{inputenc}
\usepackage{comment}
\usepackage{BOONDOX-cal}
\usepackage{ulem}
\usepackage{bm}

\begin{document}

\title{Heating rate measurement and characterization of a prototype surface-electrode trap for optical frequency metrology}

\author{Thomas Lauprêtre, Bachir Achi, Lucas Groult, Yann Kersalé, Marion Delehaye, Moustafa Abdel Hafiz and Clément Lacroûte%
\small

FEMTO-ST Institute, Université Bourgogne Franche-Comté (UBFC), CNRS, ENSMM, 26 rue de l'\' Epitaphe, 25000 Besançon, France

clement.lacroute@femto-st.fr}
%



\maketitle

\begin{abstract}
We present the characterization of a prototype surface-electrode (SE) trap  as a first step towards the realization of a compact, single-ion optical clock based on Yb$^+$. The use of a SE trap will be a key factor to benefit from clean-room fabrication techniques and technological advances made in the field of quantum information processing. We succesfully demonstrated trapping at a 500 $\bm{\upmu}$m electrodes distance and characterized our trap in terms of lifetime and heating rate. This is to our knowledge the highest distance achieved for heating rates measurements in SE traps. This simple 5-wire design realized with simple materials yields a heating rate of $\mathbf{8\times 10^3}$ phonons/s. We provide an analysis of the performances of this prototype trap for optical frequency metrology.
\end{abstract}


\section*{Introduction}

While the frame for single-ion optical clocks and trapped-ion quantum computing was set in the late XX$^{\rm{th}}$ century \cite{Dehmelt1982, Cirac1995}, experimental progress accelerated drastically during the past 15 years. Single-ion optical frequency standards have reached stunning levels of accuracy \cite{Huntemann2016, Brewer2019}, and the realization of optical time scales and optical clocks networks is becoming a reality \cite{Letargat2013, Grebing2016, Beloy2021}. On the other hand, technological progress of SE traps has allowed tremendous advances both for quantum computation \cite{Mehta2020} and for the integration of advanced functions in compact physical packages \cite{Romaszko2020}.

We are developing a compact single-ion clock based on a SE trap \cite{Lacroute2016} using the 435.5~nm quadrupole transition in $^{171}$Yb$^+$. Our target fractional frequency stability of $<10^{-14}\tau^{-1/2}$ is ten times below commercial masers performances. With robust continuous operation and high output power at $2\times435.5=871$~nm, it would constitute a perfect optical flywheel for an operational all-optical timescale. Low systematic shifts and reproducibility would further enable the use of such a clock as a mobile reference in an optical clock network, and accuracy on the order of $10^{-17}$ would be attractive for relativistic geodesy campaigns.

We have tested a simple SE geometry using a printed circuit board (PCB) trapping chip on an FR4 substrate. FR4 (Flame Retardant 4) is a fiber glass reinforced epoxy resin with low outgasing properties \cite{Rouki2003}. Several teams have developed PCB traps \cite{Brown2007, Szymanski2012, Guise2014} as a simple, low cost way to test a trapping configuration. To our knowledge, we are the first team developing such an SE trap for optical frequency metrology. Moreover, we have specifically chosen to work at a large trapping distance in order to prevent anomalous heating, which approximately scales as $1/d^4$ where $d$ is the ion to electrodes distance. The use of a prototype PCB trap has allowed us to perform a pre-characterization of the potential performances of such a setup.

In this manuscript, we first describe our single-ion trap setup, and the trapping potential experimental characterization. We then present our heating rate measurements and a theoretical estimation of the performances of a single-ion clock based on such a trap.

\section{Experimental setup}

\begin{figure}[b!]
   \includegraphics[trim={0 0cm 0cm 0},clip,width=\linewidth]{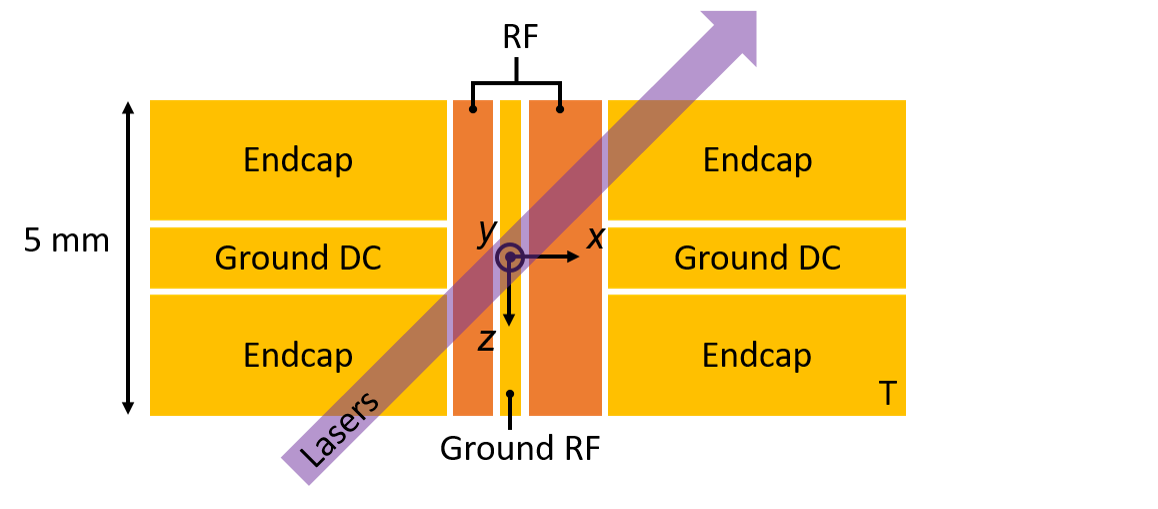}     
	\caption{Drawing of the trap electrodes (5-wire geometry). Yellow: DC electrodes. Orange: RF electrodes. Violet arrow: laser beam direction. T: connection of a bias tee outside the chamber for modulation of the corresponding DC electrode voltage.}
	\label{fig:trap}
\end{figure}


The trap used in this manuscript is a FR4 PCB circuit with a 5-wire geometry \cite{Chiaverini2005} using mini-SD in-vacuum electrical connection, described in \cite{Lacroute2016, Delehaye2018}. Fig. \ref{fig:trap} shows the electrodes layout. The RF electrodes are 630~$\upmu$m and 1200~$\upmu$m wide. With the 330~$\upmu$m wide ground RF electrode, this results in a 500~$\upmu$m trapping distance. The RF electrodes asymmetry tilts the RF pseudopotential axes $(x',y')$ in the $(x,y)$ plane \cite{Chiaverini2005}, and lifts the trapping frequencies degeneracy.

RF electrodes connected to a low-loss RF inductance form a resonant LC circuit with resonance angular frequency $\Omega_\mathrm{RF}=2\pi\times$5.7\,MHz. Peak-to-peak voltages up to $V_\mathrm{RF}=280$\,V$_\mathrm{pp}$ can be obtained at the RF electrodes, resulting in theoretical trapping frequencies of up to 350~kHz for a single $^{176}$Yb$^+$ ion (all experiments presented in this article were performed with the $^{176}$Yb isotope for simplicity).  Theoretical values are computed using an analytic model \cite{House2008}.

The trap is housed in an octogonal titanium chamber \cite{Delehaye2018}. A commercial ytterbium dispenser\footnote{Alvatec, now AlfaVakuo.} outputs the atoms, and ions are produced by a two-stage photo-ionization process \cite{Johanning2011} using an isotopically selective $399$~nm laser \cite{Laupretre2020} and the 369.5~nm laser also used for Doppler cooling. Two additional lasers at 935.5~nm and 638~nm are used to repump the ions from the $^2$D$_{3/2}$ and $^2$F$_{7/2}$ states, respectively.

All four lasers are commercial extended cavity diode lasers (ECDL)
, and are brought to the experiment by a single endlessly monomode photonic crystal fiber\footnote{NKT Photonics.} that outputs the beam at a $45^{\circ}$ angle in the $x-z$ plane shown Fig. \ref{fig:trap}. This provides a non-zero projection along all trap axes in order to cool down all three degrees of freedom \cite{Itano1982}. The cooling laser is focused on a $60~\upmu$m waist spot close to the center of the trap, providing saturation parameters of several tens with powers of just a few tens of microwatts while limiting any charging of dielectric material from UV lasers. All lasers are frequency stabilized using a commercial wavemeter \cite{Saleh2015}.

We detect the trapped ions fluorescence at 369.5 nm using a photomultiplier (PM) and a high quantum efficiency Electron-Multiplying CCD (EMCCD) camera.

\section{Trap characterization}

One of the fundamental requirements for an ion trap set-up is to keep ions long enough for an experiment to be conducted.  Trapping times up to several weeks or months are regularly reported in 3D traps \cite{Tamm2009}, and  trapping times of the order of an hour in absence of cooling lasers are now also common with SE traps \cite{McLoughlin2011}.

The background pressure is 5.5$\times10^{-10}$~mbar, maintained by a hybrid ion-getter pump. The resulting lifetime, with cooling lasers on, is about 25 min. This is comparable with some other PCB traps lifetimes \cite{Brown2007, Szymanski2012}. This lifetime is compatible with a local partial pressure of order $10^{-10}\ \rm{mbar}$ in the ion vicinity for species susceptible of molecule formations. Collisions can indeed lead to the formation of charged molecules such as YbO$^{+}$ or YbH$^{+}$ \cite{Sugiyama1995,Hoang2020}. 
It is remarkable that such a low background pressure could be achieved with a standard FR4 PCB and a non UHV-graded mini-SD slot.


We have measured the trap secular frequencies with relative uncertainty below 5$\times10^{-3}$ by modulating the DC field amplitude \cite{Nagerl1998} through a corner electrode (indicated by a ``T" on Fig. \ref{fig:trap}). 
The axial frequency is measured to be 85~kHz, in agreement with the theoretical prediction. We also find a very good agreement between theoretical predictions and experimental measurements, with radial frequencies ranging from 225 to 350 kHz for RF amplitudes between 200 and 275 V.

One of the main challenges of ion trapping experiments is the ability to minimize excess micromotion (EMM) driven by the RF potential \cite{Berkeland1998}. EMM affects clock operation through the relativistic second-order Doppler shift \cite{Ludlow2015}. Experimental imperfections that cause EMM, like residual electric fields or RF electrodes phase mismatches for instance, also lead to an increase of the DC Stark shift associated to the time averaged squared electric field sensed by the ion \cite{Berkeland1998}. It is therefore crucial to deterministically move the trapping position to the RF null and to measure the residual EMM.

When using a camera to image the ions, it is common with linear Paul traps to increase the RF peak-to-peak amplitude to observe the trap position converge towards the RF null \cite{Berkeland1998}. This method has also been used to very precisely compensate EMM with SE trap geometries \cite{Gloger2015}. We have used this method as an initial coarse tuning, by positioning a single ion such that its spot center position, focus and fluorescence level stay identical when varying the RF peak-to-peak amplitude over the entire range (100\,V to 270\,V). With a resolution of 2.7\,$\upmu$m.pixel$^{-1}$ and ion spots full width at half maximum (FWHM) of 7\,$\upmu$m, we can expect to position the ion within $\Delta u=3\ \upmu$m of the RF null in each transverse direction. The SE trap simulations indicate a corresponding micromotion amplitude upper-bound of $A_\mathrm{RF,rad}\approx370$\,nm. We can evaluate the maximum residual static electric field to be $E_\mathrm{trans}=(\sqrt{2}\Delta u)\frac{M\omega_\mathrm{rad}^2}{e}<37$\,V/m with $e$ the elementary charge.

A reduction below 10\,V/m is required for clock operation with systematic shifts at the $10^{-17}$ level \cite{Abdelhafiz2019}. In order to enhance further the micromotion compensation, standard techniques rely on photon-correlation or sideband spectroscopy \cite{Berkeland1998,Keller2015}. We rather modulate the RF voltage amplitude at the trap secular frequencies so that the ion motion is resonantly excited and scan the ion position to find the least sensitive spot \cite{Tanaka2012}.

%

The influence of the RF voltage on the trap axial position is very low, but the presence of experimental stray fields can introduce non-negligible EMM along this direction. We have measured the single ion spot width as a function of its axial position while modulating the RF amplitude at the axial secular frequency. With this method, it is possible to position the ion within $\Delta z=\pm$5\,$\upmu$m of the RF  null in the axial direction, corresponding to a residual electric field $E_z=\Delta z\frac{M\omega_\mathrm{ax}^2}{e}\leq2.6$\,V/m. According to simulations, this is also equivalent to a maximum micromotion amplitude along the axis  $A_\mathrm{RF,ax}\leq5$\,nm.

We move the ion in the transverse plane around the RF null position by using alternatively the ground RF electrode and a set of ground DC electrodes. We directly observe the PM fluorescence level detected while slowly sweeping the RF modulation frequency on a 5\,kHz range around one of the transverse secular frequencies. When the modulation frequency hits the secular frequency, the excitation of the ion motion is detected as a dip in the fluorescence spectrum. By minimizing the amplitude of this excitation alternatively in each tuning direction, the ion can be positioned to the RF null with a resolution of $\pm5\times10^{-3}$~V on the control voltages. According to simulations, this corresponds to a maximum distance from the RF null $|\Delta r|\leq\pm0.7\,\upmu$m (mostly along the $y'$ axis), with a maximum micromotion amplitude  $A_\mathrm{RF,rad}\leq60$\,nm. The maximum residual static electric field is evaluated to be $|E_\mathrm{rad}|=\Delta r\frac{M\omega_\mathrm{rad}^2}{e}\leq6.2$\,V/m in the transverse plane. This is below the required limit of 10\,V/m mentioned above.

\section{Heating rate measurements}\label{sec:heatingrate}

When using trapped ions for coherent spectroscopy and for quantum information applications, the ions vibrational state should ideally be preserved during a probing sequence even in absence of cooling. The heating rate induced by electric noise at the trap electrodes \cite{Turchette2000} is thus another fundamental parameter that needs characterization and has led to numerous works where it is quantified and minimized \cite{Brownnutt2015}. The heating rate is quantified through the measurement of the temperature change of a single trapped ion using techniques such as the Doppler recooling method \cite{Wesenberg2007,Epstein2007}, Raman spectroscopy \cite{Brownnutt2015} and the analysis of the ion spot pictures \cite{Boldin2018, Srivathsan2019}.

We have used ion spot pictures measurements and the Doppler recooling method. In both cases, we estimate the ion temperature after a controlled warm-up time where all lasers are turned off. Measuring the temperature of the ion as a function of this ``dark time'' provides an experimental value of the heating rate.

Experimentally achievable dark times are limited by the beam waist of our cooling laser at the trap position and the axial frequency of $85$\,kHz. Taking into account the 45$^{\circ}$ projection of the probe laser on the secular axis, we limit the highest measurable temperatures to 500~mK in order to neglect the influence of the spatial transverse variations of the beam intensity. This upper bound for temperature measurements also induces a limitation on the timescale of measurable Doppler recooling dynamics.

\subsection{Ion spot picture analysis}

For temperatures down to the Doppler limit in the weak binding regime, the root mean square (RMS) value $\sigma_u$ of the Gaussian position distribution of the ion is, in any of the secular axis direction $u$ \cite{Knunz2012, Rajagopal2016, Laupretre2019}:
\[
\sigma_u=\sqrt{\frac{k_\mathrm{B}T_u}{M\omega_u^2}}
\] 
with $T_u$ the temperature along the secular direction, $k_\mathrm{B}$ the Boltzmann constant and $\omega_u$ the trapping angular frequency along direction $u$. The imaged fluorescence spot represents the ion position distribution convoluted with the imaging Point Spread Function (PSF). Assuming the PSF to be Gaussian with RMS value $\sigma_\mathrm{PSF}$, the detected spot is also Gaussian with RMS value $\sigma_\mathrm{im}$ \cite{Knunz2012}:
\[
\sigma_\mathrm{im}^2=\sigma_\mathrm{PSF}^2+\sigma_u^2.
\]
A difference $\Delta T_u$ between temperatures $T_{u,1}$ and $T_{u,2}$ is easily expressed as a function of the corresponding measured spot sizes $\sigma_\mathrm{im,1}$ and $\sigma_\mathrm{im,2}$:
\[
\Delta T_u=T_{u,1}-T_{u,1}=\frac{M\omega_u^2}{k_\mathrm{B}}\left(\sigma_\mathrm{im,2}^2-\sigma_\mathrm{im,1}^2\right).
\]
The heating rate can be extracted from the measurements of the ion spot size as a function of the dark time, and results will therefore be presented as total image gaussian variance $\sigma_\mathrm{im}^2$ for simplicity an clarity.

The pictures exposure time needs to be long enough to ensure good signal-to-noise ratio (SNR) but short enough in regards of the cooling dynamics. We use the maximum available laser power at a detuning $\delta \nu_\mathrm{pic}\approx-5$\,MHz small compared to the cooling transition natural linewidth $\Gamma_{\rm{cool}}=23$\,MHz, so that the cooling dynamics stay slow according to Doppler recooling simulations, and the fluorescence level is close to its maximum while remaining on the cooling side of the transition. 
According to Doppler recooling simulations with our experimental parameters, at the end of the exposure time, the temperature should not have dropped to less than 80\% of its initial value. By measuring the ion spot size as a function of the picture exposure time, we have also checked experimentally that we cannot detect any reduction from the initial temperature with this value of 50$\,\upmu$s. As a conservative estimate, 25 \% will still be added to the heating rate results to account for the temperature reduction caused by the exposure time.

In the experimental sequence, the 370\,nm laser is kept ON only during the 50\,$\upmu$s exposure time, and shut down for a safety duration using an AOM during the camera readout. After the picture has been taken, the ion is cooled down for 50\,ms by dropping the power of the 370\,nm laser to 2.5\ I$_\mathrm{sat}$ and increasing the detuning to $\delta \nu_\mathrm{cool}\approx-20$\,MHz. These parameters are chosen experimentally as a compromise to protect the ion against high energy collisions, at the expense of a relatively high temperature at the end of this cooling period. All lasers are then shut down using mechanical shutters to let the ion warm up for a controlled dark time.

In order to free oneself from the minimum resolution determined by the aberrated PSF, a set of data has also been recorded when artificially warming up the ion for 750\,$\upmu$s before the dark time, by detuning the 370\,nm laser to $\delta \nu_\mathrm{cool}\approx+40$\,MHz with maximum available power. Under these conditions, we ensure $\sigma_u \gg \sigma_{\rm{PSF}}$.

\begin{figure}[b!]
	\centering
   \includegraphics[trim={0 0cm 0cm 0cm},clip,width=0.62\linewidth]{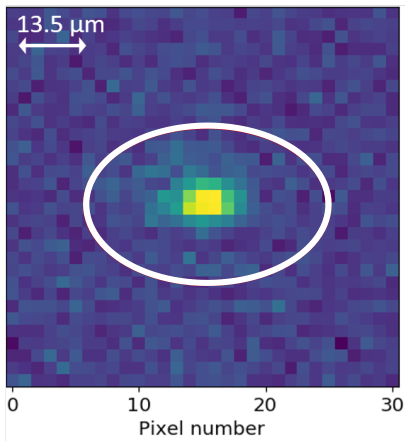}     
	\caption{Ion picture obtained from the average of $\sim$4000 sequences after a 1\,s dark time when the ion has been voluntarily warmed up at the end of the cooling period. The counts number is indicated by the color scale. The red elliptical contour represents the area outside of which the baseline plane is fitted and inside of which data are fitted.}
	\label{fig:Ion_picture}
\end{figure}

Averaged images of 4000 sequences are fitted by a 2D gaussian function inside an elliptical contour, as illustrated Fig.~\ref{fig:Ion_picture}. A baseline plane is obtained and substracted by fitting the background noise outside the ellipse to reject slow fluctuations.


\begin{figure}[b!]
\centering
  \begin{minipage}[c]{0.66\linewidth}
	\textbf{a)}\\
   \includegraphics[trim={0 0cm 0cm 0cm},clip,width=\linewidth]{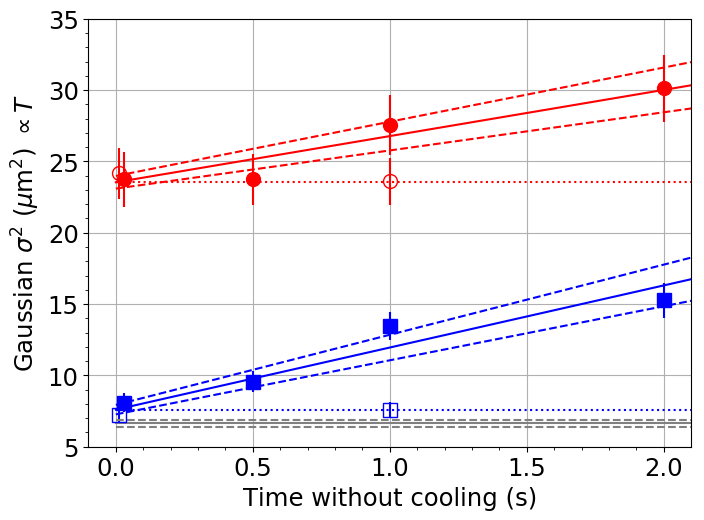}     
  \end{minipage}
	\hfill
	\begin{minipage}[c]{0.66\linewidth}
   \textbf{b)}\\
   \includegraphics[trim={0 0cm 0cm 0},clip,width=\linewidth]{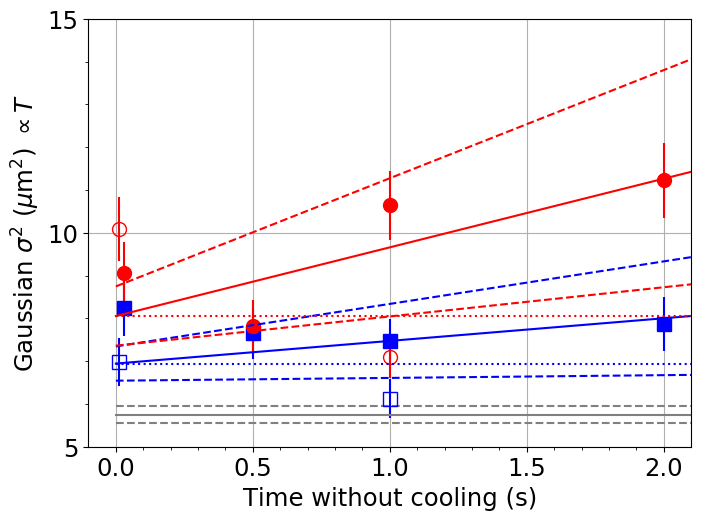}     
  \end{minipage}
	\caption{Variance of the 2D gaussian fit of the ion spot as a function of the dark time without cooling. Full blue squares are the measurements after the dark time while open blue squares are the measurements after the cooling period. Full red disks and open red circles are the corresponding measurements when the ion is voluntary warmed up after the cooling period. The gray straight line indicates the measurement for an ion in continuous cooling and detection, with its error bars indicated by the dashed gray lines. Red and blue continuous lines are the linear fit results of the data with red and blue dashed lines indicating the fit error bars. \textbf{a)} In the axial direction and \textbf{b)} in the detected transverse direction (along the $x$-axis).}
	\label{fig:HR}
\end{figure}

Fig. \ref{fig:HR} \textbf{a)} displays the final data results for the axial direction after a simple dark time sequence as full blue squares, and with an initial warm-up pulse as full red disks. The corresponding open symbols are the results of the  pictures taken just before the dark time for reference. These confirm that cooling is efficient, and are used to determine the measurements ordinate offset. A linear fit of the data returns slopes of 30$\pm$4\,mK.s$^{-1}$ and 22$\pm$4\,mK.s$^{-1}$, respectively. The heating rate is independent of the ion initial temperature within our error bars, which confirms that the measurement is not impaired by the imaging resolution.


The results for the $x$ direction are displayed on Fig. \ref{fig:HR} \textbf{b)} using the same denotation. Because of the higher secular frequency in this direction, ion spot sizes are smaller and data barely exceed the measurement resolution. Similar linear fits return 52$\pm$45\,mK.s$^{-1}$ and 156$\pm$90\,mK.s$^{-1}$, respectively, but must only be considered as a conservative upper bound.

We have checked that these values are independent of residual excess micromotion, as it could be the case in presence of spurious noise around angular frequencies $\Omega_\mathrm{RF}\pm\omega_u$ \cite{Brownnutt2015}. To do so, we have voluntary introduced a supplementary voltage on the ground DC electrodes up to 10 times our minimum resolution for compensation and in both directions. According to trap simulations, this represents a translation of the ion out of the RF null in a composite direction of the two trap axes of up to $\pm 1.2\,\upmu$m, when operating the trap at $V_\mathrm{RF}\approx 275$\,V. This corresponds to micromotion amplitudes up to $100\,$nm. In this range, no evolution of the heating rate could be detected within our error bars.

In conclusion, this method allows to conservatively estimate the heating rate to be 33$\pm$4\,mK.s$\mathbf{^{-1}}$ in the axial direction (average of the two measured slopes with added 25 \%) and provides an upper bound of 308\,mK.s$\mathbf{^{-1}}$ for the transverse direction (maximum possible value of the slope with added 25 \%).

\subsection{Doppler recooling method}

The Doppler recooling method was theoretically introduced in \cite{Wesenberg2007} side to side with the experimental results presented in \cite{Epstein2007}. The method relies on measuring the onset of the fluorescence signal when cooling lasers are shined onto a single ion. The mean dynamics of the fluorescence signal can be fitted with a simple 1D model to return a value of the temperature of the ion at the beginning of the cooling time.

We have also implemented this technique, as it is both widely used in the community and simple to setup. We used a 5 s dark time and two different detunings, and inferred the heating rate assuming a negligible initial temperature.

For $\delta \nu_\mathrm{cool}=-5$\,MHz, the data fit returns a temperature of $0.63 \pm 0.04$\,K after 5\,s without cooling, and of $0.37 \pm 0.07$\,K for $\delta \nu_\mathrm{cool}=-20$\,MHz. They correspond to heating rates of 130\,mK.s$^{-1}$ and 75\,mK.s$^{-1}$, respectively.

The discrepancy between the ion-spot analysis and the Doppler recooling method supports a non-negligible transverse heating rate. This points out that the simple Doppler recooling method is not directly applicable in our case, because the single dimensionality assumption is not verified. In this regime, the Doppler recooling method tends to an over-evaluation of the initial temperature \cite{Wesenberg2007}. The upper bound on the heating rate in the transverse direction can safely be reduced to 130\,mK.s$^{-1}$.


Our results illustrate the fact that this method becomes inappropriate when the heating rate is comparable in two directions. In this case, only an upper bound can be obtained, as the theoretical model fails. This can mistakenly be overlooked if the Doppler-recooling method is not combined with another, direction-sensitive technique.

\section{Discussion}\label{sec:clock}

Using the combined results of the two methods presented section \ref{sec:heatingrate}, we can infer the heating rate to be $33 \pm 4$\,mK.s$^{-1}$ in the axial direction and below 130\,mK.s$^{-1}$ in the transverse direction. 
In order to put our results into perspective, we calculate the electric noise spectral density \cite{Hite2012}
\[
\omega_uS_E(\omega_u)=\frac{4M\omega_u}{Q^2}k_\mathrm{B}\dot{T}_u,
\]
to obtain in the axial direction $\omega_z S_E(\omega_z)\approx1.1\times10^{-5}$\,V$^2$.m$^{-2}$ and in the transverse direction $\omega_x S_E(\omega_x)\approx1.9\times10^{-4}$\,V$^2$.m$^{-2}$. Given our $500\,\upmu$m trapping distance, these results add up rather well to the point clouds found in the literature \cite{Hite2012}. The presence of a non-negligible transverse heating rate combined with the measured trap lifetime and the observation of high-energy collisions at a 2 min$^{-1}$ rate during Doppler-recooling sequences is a strong indication of a rather high background pressure at the trap location. This could be a source of collisional heating, which will be removed in a next experiment by using better materials. 

It can be useful when investigating clock operation to use the axial and transverse secular frequencies, 85\,kHz and 350\,kHz respectively, to express the heating rate
\[
\dot{\bar{n}}_u=\frac{k_\mathrm{B}\dot{T}_u}{\hbar\omega_u}
\]
measured with this work in both directions close to 8$\times\mathbf{10^{3}}$\,phonons.s$\mathbf{^{-1}}$.
\\

This value can be used to discuss our prototype trap properties regarding a hypothetical clock operation. As a reminder, the prospect of the experiment is to use the ${}^{2}$S$_{1/2}\leftrightarrow{}^{2}$D$_{3/2}$ quadrupolar transition at 435.5\,nm in $^{171}$Yb$^{+}$ as the clock transition. The upper state ${}^{2}$D$_{3/2}$ of this transition has a lifetime of $\tau\sim$\ 52\,ms \cite{Schacht2015,Yu2000}.

During clock interrogation, the ion temperature should stay as unaffected as possible. Ideally, the heating rate $\dot{\bar{n}}$ should satisfy $\dot{\bar{n}}\tau_{c}\leq\bar{n}$, where $\tau_{c}$ is the period spent without cooling during the sequence \cite{Abdelhafiz2019}. With our upper bound estimate for the heating rate of $\dot{\bar{n}}\sim8\times10^3$\,phonons.s$^{-1}$, this would limit $\tau_{c}$ to a few ms only when probing in the transverse direction. This corresponds to a quantum-projection-noise limited fractional frequency stability of $5\times10^{-14} \tau^{-1/2}$ assuming a 1~s cycle time. To bypass this limitation, the heating rate should be reduced below $\dot{\bar{n}}<10^{3}$\,phonons.s$^{-1}$ in the transverse direction and be negligible compared to the axial heating rate we measured with our prototype chip. These values are well within reach with state-of-the-art setups.

The ion heating can also induce a loss of contrast when fringes are washed out by the wavefunction spread-out over many motional states \cite{Meekhof1996,Meekhof1996b,Wineland1998,Roos2000}.
%
This is prevented when the Lamb-Dicke criterion $\eta^2\bar{n}\ll1$ is verified, where $\bar{n}$ is the mean phonon number and $\eta^2=\left(\frac{2\pi}{\lambda}\cos\theta_\mathrm{sec}\right)^2\frac{\hbar}{2M\omega_\mathrm{sec}}$ with $\lambda$ the laser wavelength, $\theta_\mathrm{sec}$ the angle between the laser direction and the secular axis and $\omega_\mathrm{sec}$ the trap secular frequency.

The Doppler cooling limit for ytterbium is $T_\mathrm{D}=\frac{\hbar\Gamma_{\rm{cool}}}{2k_\mathrm{B}}\approx0.47$\,mK. Along the transverse directions, our trapping frequencies lead to $\eta^2\bar{n}\approx0.25$ at 0.5\,mK. The access to higher secular frequencies would enable to work even deeper in the Lamb-Dicke regime.

Table \ref{table:soa} summarizes the estimated motion-related shifts that could be observed with the current performances of our trap. First, this includes the 2$^\mathrm{nd}$ order Doppler shift from secular and micro-motion. Because all motional frequencies and the trap field frequency are much lower than the frequency of the optical transition, this also includes the static Stark shift from the time averaged squared electric field seen by the ion  \cite{Berkeland1998,Itano2000,Ludlow2015}. The spatial spread caused by the heating rate will not contribute to an increase of the EMM, neither regarding the 2$^\mathrm{nd}$ order Doppler shift nor the related static Stark shift term. The noise spectral density leading to the measured heating rate, by itself, leads to completely negligible static Stark shift.

\begin{table}
\centering
{\begin{tabular}{l|p{1.5cm}|p{1.75cm}|p{2cm}}
Shift & Prototype trap & Prototype trap advanced & State-of-the-art	materials\\
$2^{\rm{d}}$ order Doppler & $3.1\times10^{-17}$ & $3.1\times10^{-17}$ & $\approx 10^{-18}$	 \cite{Hannig2019}	\\
Stark & $3.6\times10^{-17}$ & $1.2\times10^{-17}$ & $\approx 10^{-18}$	\cite{Lange2021}	\\
Quadrupole & $10^{-15}$ & $10^{-16}$ & $\approx 10^{-17}$ \cite{Lange2021}	\\
\end{tabular}}
\caption{Summary of the motion-related shifts that woud impair our trapping geometry. See text for details.}
\label{table:soa}
\end{table}

Our trap uses a quadrupole trap geometry. For this reason, electric field gradients are intrinsically present and lead to a quadrupole shift of the clock transition \cite{Itano2000,Ludlow2015}. The RF field contribution averages to 0 at first order. The second order can be significant \cite{Yu1994}, but  cancels out when using the right Zeeman sublevels \cite{Schneider2005b}.

Table \ref{table:soa} first column shows the shifts we expect given the results of our protype trap characterization. The second column takes into account the potential use of rejection-schemes that can be implemented to reduce the Stark and quadrupole shifts \cite{Dube2005, Tamm2009, Ludlow2015}. With such schemes, a clock based on such a simple 5-wire trap geometry and inexpensive materials would exhibit trap-related shifts on the order of a few $10^{-17}$.

We can evaluate the hypothetical performances of a clock based on the same trapping geometry with state-of-the-art materials from the literature \cite{Ludlow2015, Hannig2019, Lange2021}. For such an estimation, we rely on figures measured with 3D trap geometries, where the use of a 2D trap should not induce a difference. Table \ref{table:soa} third column displays such potential values, where the trap-related shifts are at the $10^{-17}$ level or below. This is in our opinion a strong argument supporting the choice of electrodes geometry.


\section{Conclusion} 

We have developed and characterized a PCB prototype SE trap as a first test towards the development of a compact single-ion optical clock. We have succesfully trapped ions at a 500 $\upmu$m trapping distance for the electrodes, which is to our knowledge the highest distance used in a surface ion trap for heating rate measurements.

The measured lifetime of 25 min and heating rate of $8\times10^3$ phonons/s are compatible with the literature, and are most likely limited by the background pressure in the trap vicinity. Given these results, we estimate that such a simple trap would lead to a fractional frequency instability of around $5\times10^{-14} \tau^{-1/2}$ using Rabi spectroscopy, while trap-related fractional frequency shifts would be below $10^{-16}$. Such a clock would already constitute a considerable achievement.

A trap based on the simple five-wire geometry with high trapping distance is therefore compatible with optical frequency metrology. For improved performances, the use of better materials and a lower background pressure are necessary. To both reduce the heating rate and work deeper in the Lamb-Dicke regime, higher trapping frequencies ($>500$~kHz) in the transverse plane should be used, which is compatible with state-of-the-art ion trapping techniques. Under such conditions, fractional frequency stability would be improved by the higher achievable contrasts. We estimate that trap related fractional frequency shifts (2$^{\mathrm{nd}}$-order Doppler, Stark and quadrupole shifts) would be reduced to about $10^{-17}$. Further reduction would be obtained by working with the octupole transition at 467~nm.

%
%
%

\section*{Funding}

This work has been supported by the Agence Nationale de la Recherche (ANR-14-CE26-0031-01 MITICC, ANR-10-LABX-48-01 First-TF, and ANR-11-EQPX-0033 Oscillator-IMP), the Région Bourgogne Franche-Comté, the Centre National d'\' Etudes Spatiales and the EIPHI Graduate School (contract ``ANR-17-EURE-0002").

\section*{Acknowledgments}
The authors would like to thank Alexis Mosset and Fabrice Devaux for letting us borrow the EMCCD used for heating rate measurements with ion pictures. We also thank Philippe Abb\'e, Val\'erie Soumann and Yannick Gruson for their technical help and support with the trapping chip and its electrical connections. We thank Rodolphe Boudot for his thorough reading of the manuscript.

The authors thank the MIMENTO technological facility of FEMTO-ST, part of the RENATECH network, for providing technical support.


\bibliographystyle{unsrt}
\bibliography{Biblio}

\begin{thebibliography}{10}

\bibitem{Dehmelt1982}
H.~G. Dehmelt.
\newblock Monoion oscillator as potential ultimate laser frequency standard.
\newblock {\em IEEE Transactions on Instrumentation and Measurement},
  IM-31(2):83--87, 1982.

\bibitem{Cirac1995}
J.~I. Cirac and P.~Zoller.
\newblock Quantum computations with cold trapped ions.
\newblock {\em Physical Review Letters}, 74(20):4091--4094, 1995.

\bibitem{Huntemann2016}
N.~Huntemann, C.~Sanner, B.~Lipphardt, Chr. Tamm, and E.~Peik.
\newblock Single-ion atomic clock with $3\times10^{-18}$ systematic
  uncertainty.
\newblock {\em Physical Review Letters}, 116(6):063001, 2016.

\bibitem{Brewer2019}
S.~M. Brewer, A.~M. Chen, J. S.and~Hankin, E.~R. Clements, C.~W. Chou, D.~J.
  Wineland, D.~B. Hume, and D.~R. Leibrandt.
\newblock $^{27}\mathrm{Al}^+$ quantum-logic clock with a systematic
  uncertainty below $10^{-18}$.
\newblock {\em Physical Review Letters}, 123(3):033201, 2019.

\bibitem{Letargat2013}
R.~Le~Targat, L.~Lorini, Y.~Le~Coq, M.~Zawada, J.~Guéna, M.~Abgrall, M.~Gurov,
  P.~Rosenbusch, D.~G. Rovera, B.~Nagórny, R.~Gartman, P.~G. Westergaard,
  M.~E. Tobar, M.~Lours, G.~Santarelli, A.~Clairon, S.~Bize, P.~Laurent,
  P.~Lemonde, and J.~Lodewyck.
\newblock Experimental realization of an optical second with strontium lattice
  clocks.
\newblock {\em Nature Communications}, 4, 2013.

\bibitem{Grebing2016}
C.~Grebing, A.~Al-Masoudi, S.~D\"{o}rscher, S.~H\"{a}fner, V.~Gerginov,
  S.~Weyers, B.~Lipphardt, F.~Riehle, U.~Sterr, and C.~Lisdat.
\newblock Realization of a timescale with an accurate optical lattice clock.
\newblock {\em Optica}, 3(6):563--569, 2016.

\bibitem{Beloy2021}
K.~Beloy~\textit{et al.} and {Boulder Atomic Clock Optical Network (BACON)
  Collaboration}.
\newblock Frequency ratio measurements at 18-digit accuracy using an optical
  clock network.
\newblock {\em Nature}, 591(7851):564--569, 2021.

\bibitem{Mehta2020}
K.~K. Mehta, C.~Zhang, M.~Malinowski, T.-L. Nguyen, M.~Stadler, and J.~P. Home.
\newblock Integrated optical multi-ion quantum logic.
\newblock {\em Nature}, 586(7830):533--537, 2020.

\bibitem{Romaszko2020}
Z.~D. Romaszko, S.~Hong, M.~Siegele, R.~Kahan Puddy, F.~Raphaël
  Lebrun-Gallagher, S.~Weidt, and W.~K. Hensinger.
\newblock Engineering of {Microfabricated} {Ion} {Traps} and {Integration} of
  {Advanced} {On}-{Chip} {Features}.
\newblock {\em Nature Reviews Physics}, 2(6):285--299, 2020.

\bibitem{Lacroute2016}
C.~Lacro{\^{u}}te, M.~Souidi, P.-Y. Bourgeois, J.~Millo, K.~Saleh, E.~Bigler,
  R.~Boudot, V.~Giordano, and Y.~Kersal{\'{e}}.
\newblock Compact {Yb}$^+$ optical atomic clock project: design principle and
  current status.
\newblock {\em Journal of Physics: Conference Series}, 723:012025, 2016.

\bibitem{Rouki2003}
C.~Rouki, L.~Westerberg, and the CHICSi~Development Group.
\newblock Ultra-{High} {Vacuum} {Compatibility} {Measurements} of {Materials}
  for the {CHICSi} {Detector} {System}.
\newblock {\em Physica Scripta}, 2003(T104):107, January 2003.
\newblock Publisher: IOP Publishing.

\bibitem{Brown2007}
K.~R. Brown, R.~J. Clark, Jaroslaw Labaziewicz, Philip Richerme, David~R.
  Leibrandt, and Isaac~L. Chuang.
\newblock Loading and characterization of a printed-circuit-board atomic ion
  trap.
\newblock {\em Physical Review A}, 75(1):015401, January 2007.

\bibitem{Szymanski2012}
B.~Szymanski, R.~Dubessy, B.~Dubost, S.~Guibal, J.-P. Likforman, and
  L.~Guidoni.
\newblock Large two dimensional {Coulomb} crystals in a radio frequency surface
  ion trap.
\newblock {\em Applied Physics Letters}, 100(17), April 2012.
\newblock Publisher: American Institute of Physics.

\bibitem{Guise2014}
N.~D. Guise, S.~D. Fallek, H.~Hayden, C.-S. Pai, C.~Volin, K.~R. Brown, J.~T.
  Merrill, A.~W. Harter, J.~M. Amini, L.~M. Lust, K.~Muldoon, D.~Carlson, and
  J.~Budach.
\newblock In-vacuum active electronics for microfabricated ion traps.
\newblock {\em Review of Scientific Instruments}, 85(6):063101, June 2014.

\bibitem{Chiaverini2005}
J.~Chiaverini, R.~B. Blakestad, J.~Britton, J.~D. Jost, C.~Langer,
  D.~Leibfried, R.~Ozeri, and D.~J. Wineland.
\newblock Surface-electrode architecture for ion-trap quantum information
  processing.
\newblock {\em Quantum Info. Comput.}, 5(6):419–439, 2005.

\bibitem{Delehaye2018}
M.~Delehaye and C.~Lacroûte.
\newblock Single-ion, transportable optical atomic clocks.
\newblock {\em Journal of Modern Optics}, 65(5-6):622--639, 2018.

\bibitem{House2008}
M.~G. House.
\newblock Analytic model for electrostatic fields in surface-electrode ion
  traps.
\newblock {\em Physical Review A}, 78:033402, 2008.

\bibitem{Johanning2011}
M.~Johanning, A.~Braun, D.~Eiteneuer, C.~Paape, C.~Balzer, W.~Neuhauser, and
  C.~Wunderlich.
\newblock Resonance-enhanced isotope-selective photoionization of {YbI} for ion
  trap loading.
\newblock {\em Applied Physics B}, 103(2):327--338, 2011.

\bibitem{Laupretre2020}
T.~Laupr\^{e}tre, L.~Groult, B.~Achi, M.~Petersen, Y.~Kersal\'{e}, M.~Delehaye,
  and C.~Lacro\^{u}te.
\newblock Absolute frequency measurements of the
  {$^1$S$_0$$\rightarrow$$^1$P$_1$} transition in ytterbium.
\newblock {\em OSA Continuum}, 3(1):50--57, 2020.

\bibitem{Itano1982}
W.~M. Itano and D.~J. Wineland.
\newblock Laser cooling of ions stored in harmonic and {Penning} traps.
\newblock {\em Physical Review A}, 25:35--54, 1982.

\bibitem{Saleh2015}
K.~Saleh, J.~Millo, A.~Didier, Y.~Kersal\'{e}, and C.~Lacro\^{u}te.
\newblock Frequency stability of a wavelength meter and applications to laser
  frequency stabilization.
\newblock {\em Applied Optics}, 54(32):9446--9449, 2015.

\bibitem{Tamm2009}
C.~Tamm, S.~Weyers, B.~Lipphardt, and E.~Peik.
\newblock Stray-field-induced quadrupole shift and absolute frequency of the
  {688-THz $^{171}\mathrm{Yb}^{+}$} single-ion optical frequency standard.
\newblock {\em Physical Review A}, 80:043403, 2009.

\bibitem{McLoughlin2011}
J.~J. McLoughlin, A.~H. Nizamani, J.~D. Siverns, R.~C. Sterling, M.~D. Hughes,
  B.~Lekitsch, B.~Stein, S.~Weidt, and W.~K. Hensinger.
\newblock Versatile ytterbium ion trap experiment for operation of scalable
  ion-trap chips with motional heating and transition-frequency measurements.
\newblock {\em Physical Review A}, 83:013406, 2011.

\bibitem{Sugiyama1995}
K.~Sugiyama and J.~Yoda.
\newblock Disappearance of {Yb}$^{+}$ in excited states from rf trap by
  background gases.
\newblock {\em Japanese Journal of Applied Physics}, 34(Part 2, No.
  5A):L584--L586, 1995.

\bibitem{Hoang2020}
T.~M. Hoang, Y.-Y. Jau, R.~Overstreet, and P.~D.~D. Schwindt.
\newblock {${\mathrm{YbH}}^{+}$} formation in an ytterbium ion trap.
\newblock {\em Physical Review A}, 101:022705, 2020.

\bibitem{Nagerl1998}
H.~C. N\"{a}gerl, D.~Leibfried, F.~Schmidt-Kaler, J.~Eschner, and R.~Blatt.
\newblock Coherent excitation of normal modes in a string of {Ca$^+$} ions.
\newblock {\em Optics Express}, 3(2):89--96, 1998.

\bibitem{Berkeland1998}
D.~J. Berkeland, J.~D. Miller, J.~C. Bergquist, W.~M. Itano, and D.~J.
  Wineland.
\newblock Minimization of ion micromotion in a paul trap.
\newblock {\em Journal of Applied Physics}, 83(10):5025--5033, 1998.

\bibitem{Ludlow2015}
A.~D. Ludlow, M.~M. Boyd, J.~Ye, E.~Peik, and P.~O. Schmidt.
\newblock Optical atomic clocks.
\newblock {\em Review of Modern Physcis}, 87:637--701, 2015.

\bibitem{Gloger2015}
T.~F. Gloger, P.~Kaufmann, D.~Kaufmann, M.~T. Baig, T.~Collath, M.~Johanning,
  and C.~Wunderlich.
\newblock Ion-trajectory analysis for micromotion minimization and the
  measurement of small forces.
\newblock {\em Physical Review A}, 92:043421, 2015.

\bibitem{Abdelhafiz2019}
M.~Abdel-Hafiz \textit{et al.}
\newblock Guidelines for developing optical clocks with $10^{-18}$ fractional
  frequency uncertainty, 2019.
\newblock arXiv:1906.11495.

\bibitem{Keller2015}
J.~Keller, H.~L. Partner, T.~Burgermeister, and T.~E. Mehlstäubler.
\newblock Precise determination of micromotion for trapped-ion optical clocks.
\newblock {\em Journal of Applied Physics}, 118(10):104501, 2015.

\bibitem{Tanaka2012}
U.~Tanaka, K.~Masuda, Y.~Akimoto, K.~Koda, Y.~Ibaraki, and S.~Urabe.
\newblock Micromotion compensation in a surface electrode trap by parametric
  excitation of trapped ions.
\newblock {\em Applied Physics B}, 107(4):907--912, 2012.

\bibitem{Turchette2000}
Q.~A. Turchette, Kielpinski, B.~E. King, D.~Leibfried, D.~M. Meekhof, C.~J.
  Myatt, M.~A. Rowe, C.~A. Sackett, C.~S. Wood, W.~M. Itano, C.~Monroe, and
  D.~J. Wineland.
\newblock Heating of trapped ions from the quantum ground state.
\newblock {\em Physical Review A}, 61:063418, 2000.

\bibitem{Brownnutt2015}
M.~Brownnutt, M.~Kumph, P.~Rabl, and R.~Blatt.
\newblock Ion-trap measurements of electric-field noise near surfaces.
\newblock {\em Review of Modern Physics}, 87:1419--1482, 2015.

\bibitem{Wesenberg2007}
J.~H. Wesenberg, R.~J. Epstein, D.~Leibfried, R.~B. Blakestad, J.~Britton,
  J.~P. Home, W.~M. Itano, J.~D. Jost, E.~Knill, C.~Langer, R.~Ozeri,
  S.~Seidelin, and D.~J. Wineland.
\newblock Fluorescence during doppler cooling of a single trapped atom.
\newblock {\em Physical Review A}, 76:053416, 2007.

\bibitem{Epstein2007}
R.~J. Epstein, S.~Seidelin, D.~Leibfried, J.~H. Wesenberg, J.~J. Bollinger,
  J.~M. Amini, R.~B. Blakestad, J.~Britton, J.~P. Home, W.~M. Itano, J.~D.
  Jost, E.~Knill, C.~Langer, R.~Ozeri, N.~Shiga, and D.~J. Wineland.
\newblock Simplified motional heating rate measurements of trapped ions.
\newblock {\em Physical Review A}, 76:033411, 2007.

\bibitem{Boldin2018}
I.~A. Boldin, A.~Kraft, and C.~Wunderlich.
\newblock Measuring anomalous heating in a planar ion trap with variable
  ion-surface separation.
\newblock {\em Physical Review Letters}, 120:023201, 2018.

\bibitem{Srivathsan2019}
B.~Srivathsan, M.~Fischer, L.~Alber, M.~Weber, M.~Sondermann, and G.~Leuchs.
\newblock Measuring the temperature and heating rate of a single ion by
  imaging.
\newblock {\em New Journal of Physics}, 21(11):113014, 2019.

\bibitem{Knunz2012}
S.~Kn\"unz, M.~Herrmann, V.~Batteiger, G.~Saathoff, T.~W. H\"ansch, and Th.
  Udem.
\newblock Sub-millikelvin spatial thermometry of a single {Doppler}-cooled ion
  in a {Paul} trap.
\newblock {\em Physical Review A}, 85:023427, 2012.

\bibitem{Rajagopal2016}
V.~Rajagopal, J.~P. Marler, M.~G. Kokish, and B.~C. Odom.
\newblock Trapped ion chain thermometry and mass spectrometry through imaging.
\newblock {\em European Journal of Mass Spectrometry}, 22(1):1--7, 2016.

\bibitem{Laupretre2019}
T.~Laupr\^etre, R.~B. Linnet, I.~D. Leroux, H.~Landa, A.~Dantan, and
  M.~Drewsen.
\newblock Controlling the potential landscape and normal modes of ion coulomb
  crystals by a standing-wave optical potential.
\newblock {\em Physical Review A}, 99:031401, 2019.

\bibitem{Hite2012}
D.~A. Hite, Y.~Colombe, A.~C. Wilson, K.~R. Brown, U.~Warring, R.~J\"ordens,
  J.~D. Jost, K.~S. McKay, D.~P. Pappas, D.~Leibfried, and D.~J. Wineland.
\newblock 100-fold reduction of electric-field noise in an ion trap cleaned
  with in situ argon-ion-beam bombardment.
\newblock {\em Physical Review Letters}, 109:103001, 2012.

\bibitem{Schacht2015}
M.~Schacht, J.~R. Danielson, S.~Rahaman, J.~R. Torgerson, J.~Zhang, and M.~M.
  Schauer.
\newblock {$^{171}$Yb$^+$ $^5$D$_{3/2}$ hyperfine state detection and F = 2
  lifetime}.
\newblock {\em Journal of Physics B: Atomic, Molecular and Optical Physics},
  48(6):065003, 2015.

\bibitem{Yu2000}
N.~Yu and L.~Maleki.
\newblock Lifetime measurements of the ${4f}^{14}5d$ metastable states in
  single ytterbium ions.
\newblock {\em Physical Review A}, 61:022507, 2000.

\bibitem{Meekhof1996}
D.~M. Meekhof, C.~Monroe, B.~E. King, W.~M. Itano, and D.~J. Wineland.
\newblock Generation of nonclassical motional states of a trapped atom.
\newblock {\em Physical Review Letters}, 76:1796--1799, 1996.

\bibitem{Meekhof1996b}
D.~M. Meekhof, C.~Monroe, B.~E. King, W.~M. Itano, and D.~J. Wineland.
\newblock Generation of nonclassical motional states of a trapped atom.
\newblock {\em Physical Review Letters}, 77:2346--2346, 1996.

\bibitem{Wineland1998}
D.~J. Wineland, C.~Monroe, W.~M. Itano, D.~Leibfried, B.~E. King, and D.~M.
  Meekhof.
\newblock Experimental issues in coherent quantum-state manipulation of trapped
  atomic ions.
\newblock {\em Journal of research of the National Institute of Standards and
  Technology}, 103(3):259, 1998.

\bibitem{Roos2000}
C.~F. Roos.
\newblock {\em Controlling the quantum state of trapped ions}.
\newblock PhD thesis, Universität Innsbruck, Fakultät der
  Leopold-Franzens-Universität Innsbruck, Innsbruck, 2000.

\bibitem{Itano2000}
W.~M. Itano.
\newblock External-field shifts of the {$^{199}$Hg$^+$} optical frequency
  standard.
\newblock {\em Journal of research of the National Institute of Standards and
  Technology}, 105(27551639):829--837, 2000.

\bibitem{Hannig2019}
S.~Hannig, L.~Pelzer, N.~Scharnhorst, J.~Kramer, M.~Stepanova, Z.~T. Xu,
  N.~Spethmann, I.~D. Leroux, T.~E. Mehlstäubler, and P.~O. Schmidt.
\newblock Towards a transportable aluminium ion quantum logic optical clock.
\newblock {\em Review of Scientific Instruments}, 90(5):053204, 2019.

\bibitem{Lange2021}
R.~Lange, N.~Huntemann, J.~M. Rahm, C.~Sanner, H.~Shao, B.~Lipphardt, C.~Tamm,
  S.~Weyers, and E.~Peik.
\newblock Improved limits for violations of local position invariance from
  atomic clock comparisons.
\newblock {\em Physical Review Letters}, 126(1):011102, 2021.

\bibitem{Yu1994}
N.~Yu, X.~Zhao, H.~Dehmelt, and W.~Nagourney.
\newblock Stark shift of a single barium ion and potential application to
  zero-point confinement in a rf trap.
\newblock {\em Physical Review A}, 50(3):2738--2741, 1994.

\bibitem{Schneider2005b}
T.~Schneider, E.~Peik, and Chr. Tamm.
\newblock Sub-hertz optical frequency comparisons between two trapped
  {$^{171}\mathrm{Yb}^{+}$} ions.
\newblock {\em Physical Review Letters}, 94:230801, 2005.

\bibitem{Dube2005}
P.~Dubé, A.~A. Madej, J.~E. Bernard, L.~Marmet, J.-S. Boulanger, and S.~Cundy.
\newblock Electric quadrupole shift cancellation in single-ion optical
  frequency standards.
\newblock {\em Physical Review Letters}, 95(3):033001, 2005.

\end{thebibliography}






\end{document}